\documentclass[runningheads]{llncs}
\usepackage{graphicx}
\usepackage{amsfonts}
\usepackage{amsmath}
\usepackage{bbm}
\usepackage{graphicx}
\usepackage{hyperref}
\usepackage{wrapfig}
\usepackage{appendix}
\usepackage{cleveref}
\usepackage{subcaption}

\newcommand{\Ucal}{{\mathcal U}}
\newcommand{\Hcal}{{\mathcal H}}
\newcommand{\Tcal}{{\mathcal T}}

\newcommand{\E}{{\mathbb E}}
\newcommand{\Pa}{{\mathbb P}}

\newcommand{\cExp}[2]{\E \left[{#1} \mid {#2} \right]}

\newcommand{\br}{{\mathcal{A}}}

\hypersetup{colorlinks=true,
	linkcolor=blue,
	citecolor=blue,
	urlcolor=blue}

\begin{document}
\title{A Game-Theoretic Analysis of Cross-Ledger
Swaps with Packetized Payments}

\titlerunning{Game-Theoretic Analysis of Packetized Payments}

\author{
Alevtina Dubovitskaya\inst{1,2}\orcidID{0000-0002-9669-1250} \and
Damien Ackerer\inst{3}\orcidID{0000-0002-3408-2185} \and
Jiahua Xu\inst{4,5}\orcidID{0000-0002-3993-5263}}
\authorrunning{Dubovitskaya et al.}

\institute{%
Lucerne University of Applied Sciences and Arts \\
\email{alevtina.dubovitskaya@hslu.ch} 
\and
Swisscom 
\and
Covario
\\
\email{damien.ackerer@covar.io}
\and
UCL Centre for Blockchain Technologies \\
\email{jiahua.xu@ucl.ac.uk} 
\and
École Polytechnique Fédérale de Lausanne (EPFL)
} 
\maketitle              

\begin{abstract}
We propose a game-theoretic framework to study the outcomes of packetized payments, a cross-ledger transaction protocol, with strategic and possibly malicious agents. 
We derive the transaction failure rate and demonstrate that without disciplinary mechanisms, packetized payments are likely to be incomplete.
Our analysis suggests that collateral deposits can prevent malicious agents from taking advantage of the protocol. We further infer that the deposit amount should depend on the underlying asset price volatility or that it should be dynamically adjusted as the price changes.

\end{abstract}

\section{Introduction}

\subsection{Background}

\subsubsection{HTLCs}

Hashed Time Lock Contracts (HTLCs) have been recently proposed~\cite{herlihy2018atomic} to achieve atomicity of a cross-ledger transaction without any connections between the ledgers, and are often employed in decentralized exchanges (DEX)\footnote{Cross-ledger DEX protocols are not to be confused with DEX protocols operated within one chain, such as automated market makers (AMM) on Ethereum \cite{xu2021dexAmm%
}.} to complete peer-to-peer exchange \cite{komodo,decreedGit}. 

An atomic swap with HTLCs starts with one transactional agent, say Alice, randomly generating a secret key. Alice then locks her asset in an HTLC that will transfer the asset to her counterparty, say Bob, upon verification of the secret key. Bob subsequently locks his asset in an HTLC that will transfer the asset to Alice upon verification of the same secret key. The swap completes when Alice unlocks Bob's asset with the secret key generated by herself, which simultaneously exposes the secret key, allowing Bob to also unlock Alice's asset. Should Alice fail to unlock Bob's asset with her secret key, the two HTLCs will respectively send the locked asset to their original owners when the time locks expire. 

One problem of HTLCs is that they create a free option for Alice, who can ultimately choose when and whether or not to expose the secret key, thus delaying the completion of the swap or even causing it to fail. 
Bob also has the option not to lock his asset, which leads to the blocking of Alice's funds for nothing~\cite{Xu2020}.

\subsubsection{Packetized payments (PP)}

Robinson~\cite{Robinson2019} underlines the aforementioned problems associated with HTLCs, and proposes an alternative approach for cross-ledger atomic swaps: packetized payments (PP). Originally developed as part of the Interledger Protocol~\cite{Interledger2020} named {\it Hyperledger Quilt},\footnote{\url{https://github.com/hyperledger/quilt}} a cross-ledger swap with packetized payments is conducted with a series of alternating transactions. 

\begin{figure}[tb]
\centering
\begin{subfigure}[b]{0.29\textwidth}
\includegraphics[width = \textwidth]{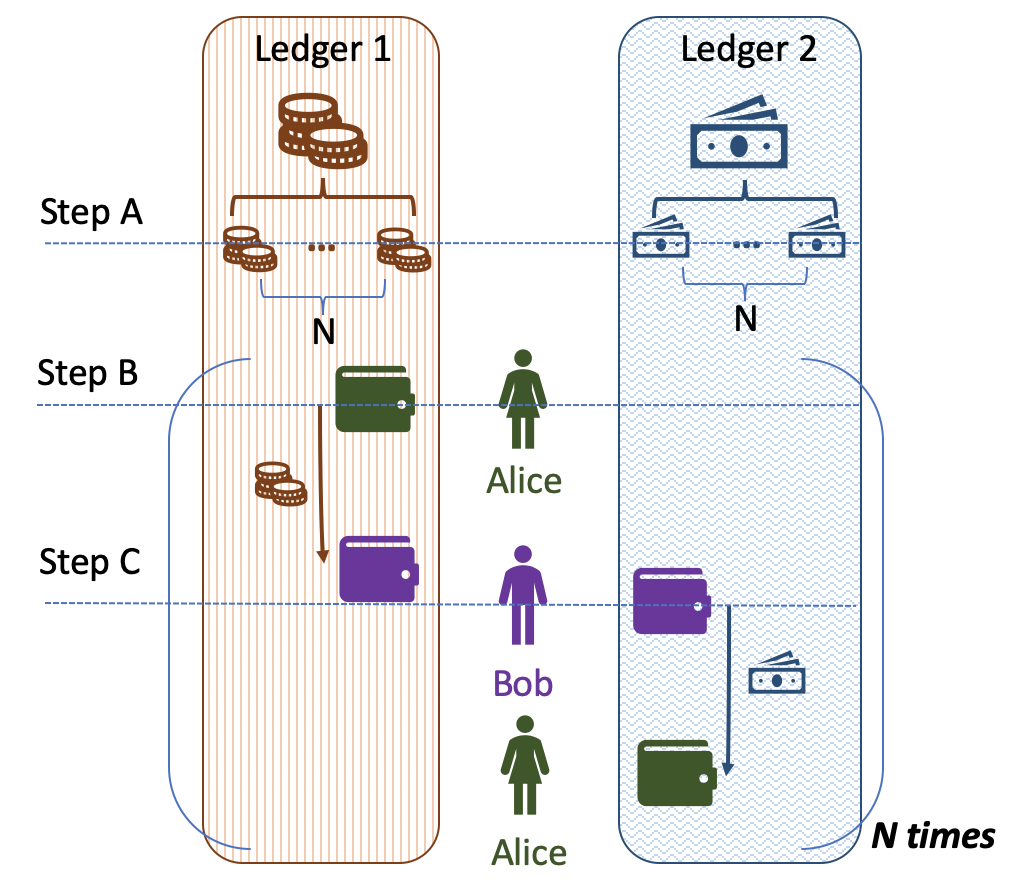}
\caption{Generic PP}
\label{fig:pp-f1}
\end{subfigure}
\hfill
\begin{subfigure}[b]{0.7\textwidth}
	   \includegraphics[width=0.49\textwidth]{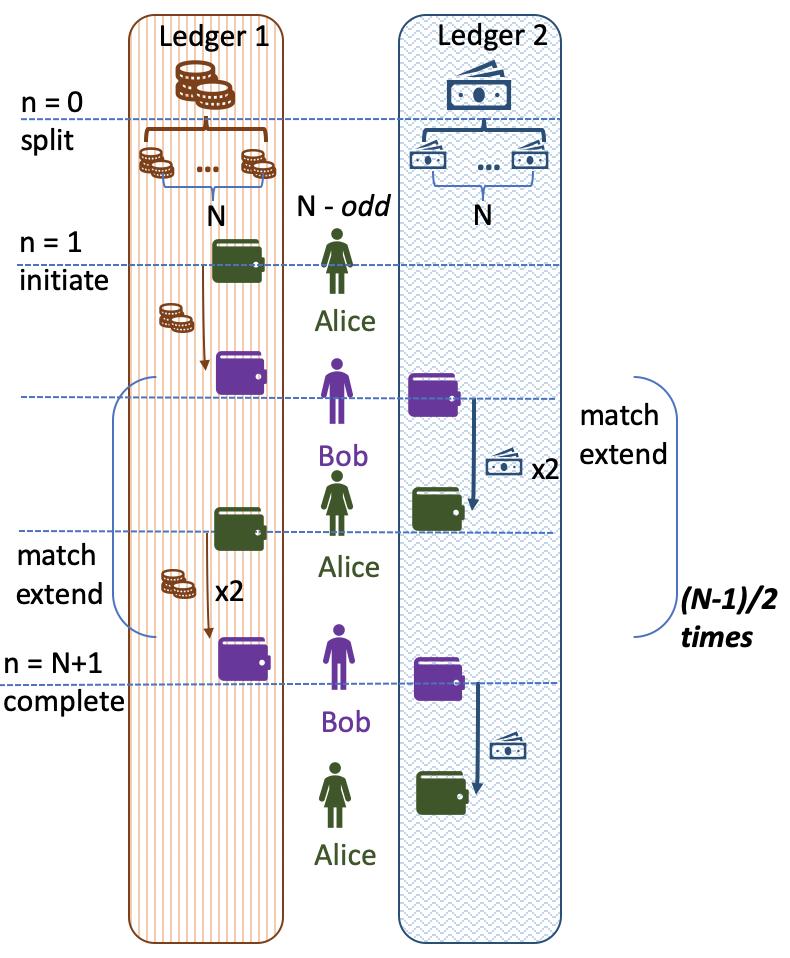}~
	   \includegraphics[width=0.49\textwidth]{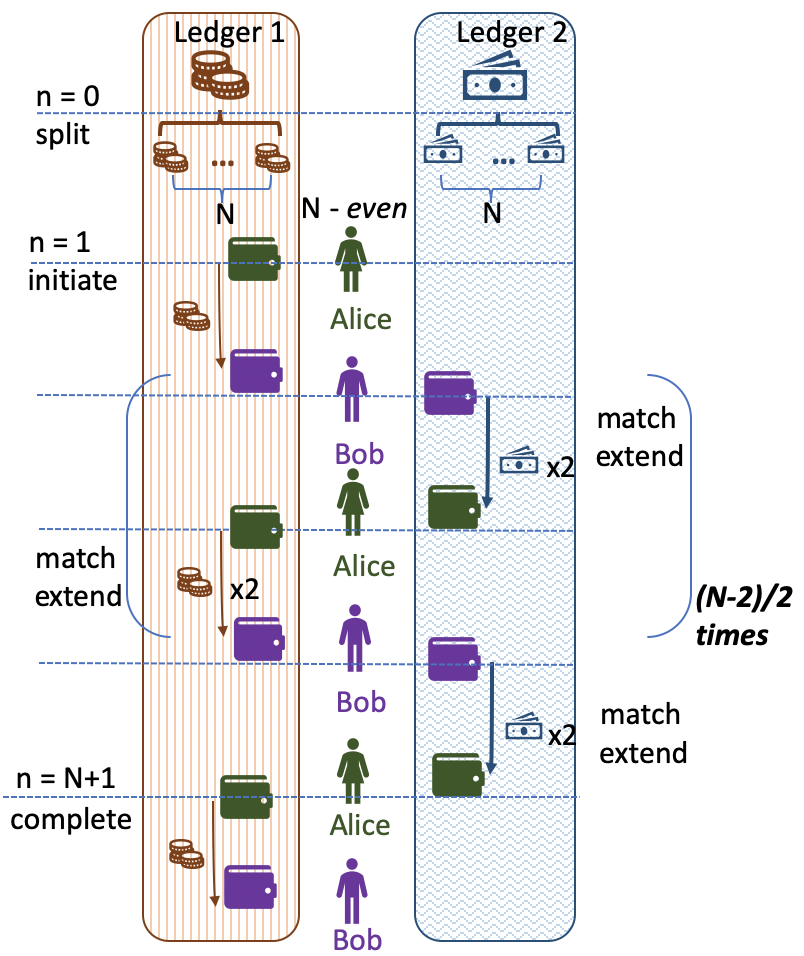}
\caption{Match-and-extend PP with odd (left) and even (right) $N$}
\label{fig:pp-fB2}
\end{subfigure}
\caption{Packetized payment (PP) schemas.}
\end{figure}

First, the total asset amounts to be traded are split into $N$ ``economically-insignificant'' amounts (\Cref{fig:pp-f1} Step A). Next, these small portions of assets will be sent on one and then on another blockchain sequentially: Steps B and C are to be repeated $N$ times in order to complete the transaction. 

Note that, at each iteration, the protocol may require the agent to match and extend the previous transfer such that the agents are alternately exposed to counterparty risk (\Cref{fig:pp-fB2}).
Otherwise, the payment initiator would have to agree to always bear the risk of abandonment from the other agent.
If one agent behaves maliciously and does not execute the transfer when it is his or her turn, the counterparty loses \emph{only} a fraction of the asset they would be willing to trade.  
Therefore, PP caps the amount of assets that can be lost at a fraction of the asset determined at Step A and prevents the whole amount of assets from being blocked for a long period of time.
It also prevents a potential loss of the whole amount of assets, while requiring only simple transfer transactions.

\subsection{Contribution}

Our framework builds on finite extensive-form games with imperfect information \cite{osborne1994course}, where the only known unknown information is the counterparty's type, which can be either honest or malicious. We study agents' strategies and derive preference parameter conditions consistent with their actions.
We also derive the transaction failure rate as a function of the percentages of honest and malicious agents.

We show that in a swap game with packetized payments, it is impossible to enforce malicious agents to complete the transaction without an additional disciplinary mechanism.
We illustrate that the ``biased'' preferences of agents for completed transactions have to be economically large, which motivates the necessity of alternative contracting mechanisms such as collateral deposits.
Still, we infer that the initial collateral amount should depend on the asset price volatility, or that it should be dynamically adjusted as the asset price fluctuates. 
As the first cross-chain packetized payment protocol {\it Hyperledger Quilt} is yet to be launched and empirical evidence is absent, our work provides the first simulation result that can facilitate further development of the protocol.

We focus on packetized payments, yet our approach can be extended to other cross-ledger transaction protocols. 

\section{A game-theoretic analysis}\label{sec:game}

\subsection{Framework} \label{sec:framework}

Two agents, Alice and Bob, or $a$ and $b$, want to exchange one unit of asset 1, say one Altcoin, from $a$ for some units of asset 2, say Tether (USDT), from $b$.
We assume that asset 2 is the reference asset in which the agents value their goods.
We denote $P_t$ the time-$t$ price of asset 1 expressed in units of asset 2, for example the price of one Altcoin in USDT.
We assume for simplicity that there is no interest rate or coin staking, meaning that the asset quantities do not increase by themselves whenever locked in a special wallet or account.
Therefore, only the price of asset 1 is stochastic in our framework.

There are three possible times $t$ at which the agents may take actions: 0, 1, and 2. 
The price dynamics of asset 1 is given by 
\begin{equation} \label{eq:price}
P_{t}=P_{t-1} \pm \delta
\end{equation}
for $t=1,2$ with equal probability of up and down moves, for some initial price $P_0>0$ and some constant $\delta>0$ such that $\delta \le \frac{P_0}{2}$ so that the price remains non-negative during the game. 
Note that the asset price is a martingale, that is the expected value of next period's price is equal to the current price, $\cExp{P_{t}}{P_{t-1}}=P_{t-1}$ for $t=1,2$.

There are three types of actions that the agents may take: continue $c$, wait $w$, and stop $s$.
If an agent plays $s$ then the game is over and the transaction fails.
If an agent plays $w$ then one time period passes and the price changes.
If an agent plays $c$ then either it is the other agent's turn, or the transaction is completed.
The agents take actions sequentially and the set of possible actions at a particular instant depends on the history of previous actions.

We assume that the agents are strategic and aim to maximize their interests which is a function of two terms: the financial profit resulting from the asset price change, and the transaction success.
Indeed, transaction failures typically have a negative economic impact on agents by delaying further trade actions, and increasing the exposure to price risk. 
We assume that there are two types of agents: the honest or high type $h$, and the malicious or low type $l$.
We formalize the two types in the following definition.

\begin{definition}[Agent types]
An agent of type $h$, namely \emph{honest}, always chooses to play continue $c$. 
An agent of type $l$, namely \emph{malicious}, satisfies the parameter condition $\alpha_{i,l}=0$ for $i=a,b$.
\end{definition}

We model the agent incentives using a utility function as follows:
\begin{equation} \label{eq:utility_func}
\Ucal(i,j) = \alpha_{i,j} X + \beta_i X Y
\end{equation}
for any agent $i\in\{a,b\}$ of type $j \in\{h,l\}$, and where $X=1$ indicates transaction success and $X=-1$ transaction failure, and $Y$ is the profit and loss resulting from the asset price change and transfer.
The constant $\alpha_{i,j}\ge 0$ measures the extent to which an agent is willing to complete the transaction.
For example, if $\alpha_{i,j}$ is large then the agent will most likely prefer to complete the transaction despite an adverse price change.
We set $\beta_b=1$ and $\beta_a=-1$ modeling the agent's opposite exposures to price changes.
Note that if the transaction fails, that is $X=-1$, then Alice is positively exposed to $Y$ because asset 1 was not transferred to Bob as $\beta_a X=1$ in this case.

In~\Cref{sec:pkt}, we derive the optimal strategy of the malicious agent, and the conditions on $\alpha_{i,h}$ such that an agent is {\it willingly honest}.

We denote $\mu_i$ the fraction of honest agents $i$ and, thus, $1-\mu_i$ the fraction of malicious agents $i$ for $i\in\{a,b\}$.
The agents meet at random, and each does not know whether the other agent is malicious or not.
Furthermore, the agents have full information about their environments. We write $\cExp{\mathcal{X}}{\mathcal{Y}}$ the expected value of the variable $\mathcal{X}$ given the history of actions and other possible refinements $\mathcal{Y}$. We write $\Tcal(i)$ the type of agent $i$, for example $\Tcal_a=h$ means that Alice is honest. 
We denote $\br(j,\Hcal)$ the best response, or action taken, by an agent of type $j$ following the history $\Hcal$, which is defined as the action maximizing their expected utility.



Note that Alice and Bob must take into account the likelihood that they can be trading with either a malicious or an honest agent.
For example, the expected utility of a type $j$ Bob conditioned on the history of actions $\Hcal$ is given by
\begin{align*}
\cExp{\Ucal(b,j)}{\Hcal}  = \, & \mu_a  \cExp{\Ucal(b,j)}{\Hcal, \Tcal_a=h} \\
 & + (1-\mu_a)  \cExp{\Ucal(b,j)}{\Hcal, \Tcal_a=l}
\end{align*}
where $\cExp{\Ucal(b,j)}{\Hcal, \Tcal_a=l}$ denotes the expected utility of type $j$ Bob under the assumption that Alice is malicious, and so on.

We use brackets to denote the history of actions, for examples $\{\emptyset\}$ for no action taken and $\{c,w,c\}$ for continue--wait--continue actions.
Which agent played a particular action and whose turn it is to play next will be clear from the game descriptions. Notations are summarized in~\Cref{tab:notations}.


\begin{table}[tp]
\begin{center}
\caption{Summary of notations \label{tab:notations}}
\begin{tabular}{c|l}
Notation & Description \\
\hline
$a$ and $b$ & Alice and Bob\\
$h$ and $l$ & honest and malicious\\
$c$, $w$, and $s$ & actions: continue, wait, and stop \\
$\Tcal_i$ & agent $i$ type \\
$\mu_i$ & honest agent $i$ percentage, $\Pa[\Tcal_i=h]$ \\
$\br(j,\Hcal)$ & agent type $j$ action after $\Hcal$ \\
$X$ & swap success (1) or failure (-1) \\
$Y$ & financial profit and loss \\
$\alpha_{i,j}$ & agent preference parameter for swap success \\
$\beta_i$ & $\beta_a=-1$ and $\beta_b=1$ indicate the asset price exposure direction \\
$P_t$ & time-$t$ price of asset 1 denominated in asset 2 \\
$\delta$ & one-period price change of asset 1 denominated in asset 2 \\
\end{tabular}
\end{center}
\end{table}

\subsection{A short packetized payment game} \label{sec:pkt}

\begin{figure}[tb]
\centering
\begin{subfigure}[b]{0.45\linewidth}
\includegraphics[width = \textwidth]{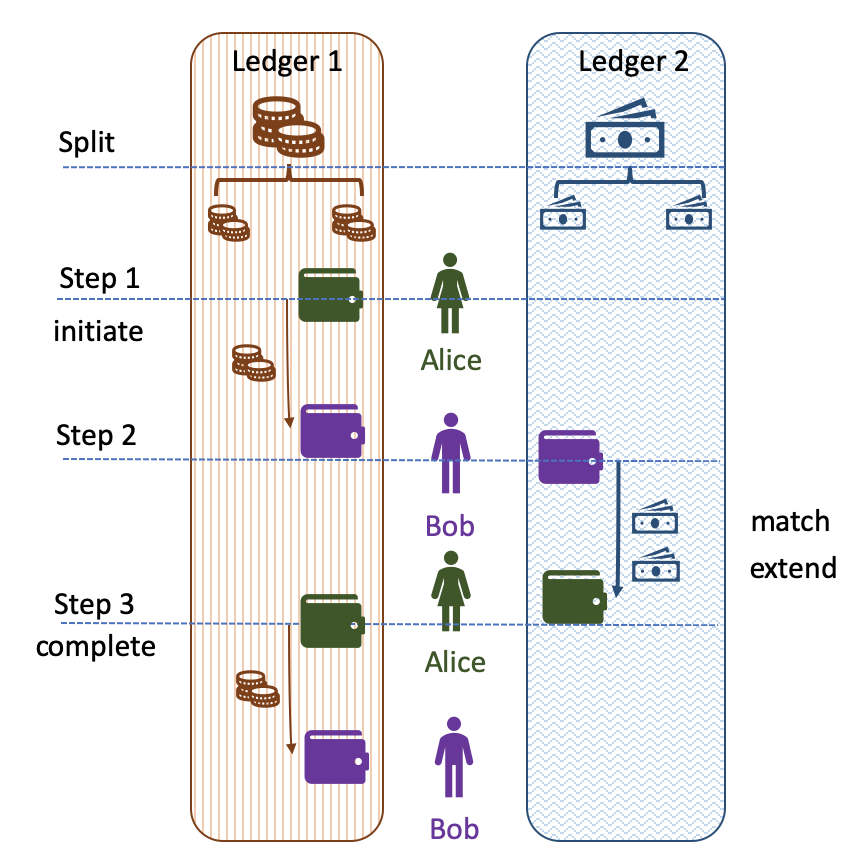}
\caption{3-step PP schema}
\label{fig:pp-fB3}
\end{subfigure}
\hfill
\begin{subfigure}[b]{0.45\linewidth}
\includegraphics[width = \textwidth]{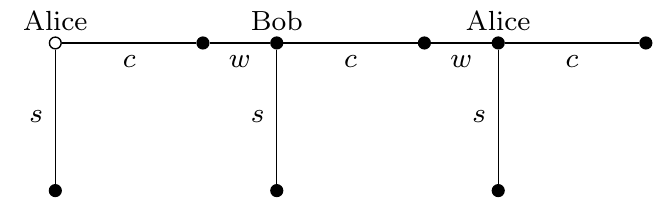}
\caption{Sequence of actions ($\circ$: root node)}
\label{fig:packet}
\end{subfigure}
\caption{Match-and-extend PP in 3 steps}
\end{figure}

Packetized payments split the transaction into small transfers where each agent exposed herself or himself to a one-way transfer alternately.
At any point in time, one agent may decide not to transfer furthermore and stop the transaction.
As a consequence, the variable $Y$ depends on the exit time and is given by 
\begin{equation} \label{eq:packet}
Y_n = \begin{cases}
        0 &  n= 0, \\ 
      \frac{n}{N} P_{t_n} - \frac{n-1}{N} P_0 & n \text{ is odd and } 0 < n \leq N \\
      \frac{n-1}{N} P_{t_n} - \frac{n}{N} P_0  & n \text{ is even and } 0 < n \leq N \\
      P_{t_n} - P_0 & n = N+1
    \end{cases}     
\end{equation}
where the subscript $n$ indicates the current step of the transaction, $\frac{1}{N}$ is the granularity amount of the PP, and $t_n$ indicates the time at which the $n$-th step takes place.

For clarity of exposition, we study a swap performed in $3$ payment transactions in total so that $t_n=n$ for $n=0,1,2$ (see \Cref{fig:pp-fB3}). 
Still, this setup is sufficient to illustrate the functioning of packetized payments.



The sequence of actions for the packetized payment game is described in \Cref{fig:packet}.
In summary, Alice transfers half of the asset to Bob, then Bob makes the whole $P_0$ payment, and finally Alice transfers the remaining half of the asset.
The agents can only decide to continue $c$ or stop $s$.
However, when an agent continues, i.e. plays $c$, we assume that the transaction also waits, i.e. plays $w$, for a short time period over which the asset price changes.
Note that the game resembles the \emph{centipede game} from \cite{rosenthal1981}, however there are important differences: the agents have partial information, and the payoffs are stochastic.
Indeed, each agent does not know the other agent type, honest or malicious, and the payoff they get depends on the asset price which is stochastic.
%

\begin{remark}
We assumed that Bob matches Alice's payment of $\frac{1}{N}$ and sends an additional $\frac{1}{N}$ payment at the same time (match and extend).
This is a fairer mechanism as the agents alternately expose themselves to a loss of $\frac{1}{N}$.
Indeed, the alternative would be to let Bob only match while Alice initiates all the payments.
This would however result in Alice being the only one exposed to counterparty risk and would require $2N$ transfers instead of $N+1$.
\end{remark}

The first and striking result is that malicious agents, either Bob or Alice, will never complete the transaction.
Indeed, there is no incentive for an agent who only cares about its financial profit to complete the transaction, as shown in \Cref{pkt:mal}. All proofs can be found in Appendix.
\begin{proposition}[Malicious Alice and Bob]\label{pkt:mal}
We have that $\br(l,\{c,w\})=s$ and $\br(l,\{c,w,c,w\})=s$.
\end{proposition}
From this result we can also infer the percentage of failed transactions.
\begin{corollary}[Transaction failure probability]\label{pkt:pfail}
Assuming that both malicious and honest agents participate in the transaction, then the percentage of incomplete transactions is $1 - \mu_b\mu_a$.
\end{corollary}
As it is always best for the malicious type to stop, the transaction will only be completed if the two agents are honest.
We derive a necessary condition so that Bob is honest and continues the transaction.
\begin{proposition}[Honest Bob]\label{pkt:bhonest}
Assume that Alice of either type plays $c$ at the initial step.
Then Bob is honest if and only if
$$
\mu_a > \frac{2P_0}{4\alpha_{b,h} + P_0 - \delta}
$$
\end{proposition}
Note that this result holds only if Bob did not update his prior that Alice is honest with probability $\mu_a$, which is the case when both Alice types play $c$ at step 0.
We observe that if the price is more volatile, meaning larger $\delta$, then a larger fraction of honest $a$ agents is required.
Interestingly, even with no price movement $\delta=0$ and only honest $a$ agents, we see that Bob's preference parameter for a completed transaction must be fairly large in value, $\alpha_{b,h}>\frac{P_0}{4}$.

We now derive conditions necessary for Alice to be honest.
\begin{proposition}[Honest Alice]\label{pkt:ahonest}
Alice is honest if and only if
\begin{equation}\label{eq:ahonest}
\alpha_{a,h} > \frac{P_0+2\delta}{4}
\quad \text{and} \quad
\mu_b > \frac{P_0}{4\alpha_{a,h} + P_0}.
\end{equation}
\end{proposition}
We see that the conditions for Alice to be honest are more stringent than for Bob even in a setup without any malicious agents, that is when $\mu_a=\mu_b=1$.
Notably if the percentage of honest Bob becomes very small, $\mu_b\rightarrow0$, then there cannot exist any honest Alice unless $\alpha_{a,h}\rightarrow+\infty$.

\section{Discussion} \label{sec:discussion}

Our game-theoretic analysis shows that there is no incentive for malicious agents to complete a PP transaction.
As a consequence, the transaction failure rate should be large and the economic incentive for agents to behave honestly would need to be enormous.
Importantly,
a malicious agent can enter multiple PP transactions in parallel with different counterparties for larger profits.

From a practical perspective, PPs are relatively simple, but require many transfers, whose total cost is therefore uncertain.
Lightning networks~\cite{poon2016bitcoin} can be employed for micropayments needed for packetized payments without incurring high transaction fees. 
However, this reintroduces the problem of the assets being locked and, in this case, in the form of collateral deposit on the escrow accounts of each agent on each blockchain: Alice and Bob will need to create two micropayment channels, one on top of each blockchain, and lock the collateral on each channel.
In addition, if an honest agent does not receive a payment from a counterparty, and is willing to close the micropayment channel, the funds on the escrow account will be blocked for a certain blockchain-specific period of time~\cite{luu2016secure}. It is also worth noticing that there may be small delays between transfers for network validation, which in turn lead to price fluctuations, as described in the PP game. Importantly, PPs cannot be used to exchange non-fungible assets such as CryptoKitties\footnote{\url{https://www.cryptokitties.co/}} or ``digital twins'' of physical goods \cite{Robinson2019}.

\subsection{Collateral deposit}

Using collateral deposits to reduce the risk of agents exposed to adverse behavior of other agents is not new. 
For example, applying collateral to disincentivize aborting a fair exchange has been discussed in works on rational fair exchange~\cite{Syverson1998}.
Zamyatin et al. \cite{Zamyatin} suggest using collateral at least equal to the assets locked on the blockchain for a trade. 
They also propose overcollateralization and a liquidation mechanism to mitigate extreme price fluctuations for short- and long-term cross-ledger transactions. 
While this ensures that economically rational agents have no incentive to misbehave, a disadvantage of this solution is that if an agent would like to transfer all their assets of one kind, they will be obliged to execute multiple transactions, each with an amount (approximately) equal to a half of the amount of the asset they currently possess.

Based on the proposed game-theoretic model, it can be shown that a marginal amount of collateral is sufficient to prevent agents from behaving maliciously. 
We modify the frameworks of~\Cref{sec:game} so as to require agents to place collateral that will be lost if they exit the transaction without completing it.
The following Proposition shows that this extinguishes malicious behaviors in our framework.

\begin{proposition}[PP with collateral] \label{pkt:collateral}
Assume that Bob places a collateral larger than $\frac{P_0+\delta}{2}$, and Alice places a collateral larger than $\frac{P_0+2\delta}{2}$, then it is optimal for malicious agents to continue the transaction in the packetized payment game described in~\Cref{sec:pkt}.
\end{proposition}

Two relevant observations can be made for real-world applications.
First, both expressions for the minimum collateral requirement involve the term $\delta$ which suggests that collateral demand should be a function of the asset price volatility, which is known to be time-varying.
Second, for the packetized payment, the initial collateral involves the fractional transfer value $\frac{P_0}{2}$, or $\frac{P_{t_n}}{N}$ in general, which suggests that the collateral requirement can be small, with $N$ large, but should be adjusted dynamically as the asset price changes.
Indeed, the price can vary up or down to $P_0\pm N\delta$ in extreme scenarios, but will in general fluctuate significantly less.

\subsection{Reputation mechanism} 
We have always assumed that an agent cannot predict the strategy of their counterparty ex-ante as the agent types, malicious or honest, are not observable. 
However, in reality, if an agent trades regularly with another agent that it can identify, or if an agent has some information on the previous behavior of another agent, then a self-selected agent matching can occur instead of a random one. 

In principle, as all the transactions executed on a ledger can generally be seen, the transaction history of an agent can be analyzed to build his or her reputation. 
However, computation of such reputation value is problematic in case of permissionless blockchains for several reasons. 
First, an agent can create multiple accounts and attempt to preserve their anonymity. 
Even though de-anonymization is possible \cite{biryukov2014deanonymisation}, one cannot guarantee a perfect mapping between one user and all his or her transactions, in the case of multiple accounts. 
Second, it may not always be possible to distinguish a cross-ledger transaction from a single-chain transaction. 
However, if these two challenges are addressed, thanks to the book-keeping property~\cite{Ibanez2020c} and immutability of a ledger, using a reputation mechanism can complement existing protocols.

\section{Conclusion and future work} \label{sec:conclusion}

We introduce a game-theoretic approach to model agent behaviors in cross-ledger transactions with packetized payments. 
We derive conditions for agents to behave honestly or maliciously, as well as different measures of economic and transaction success.
We propose to dynamically compute and adjust the collateral amounts in order to enforce honest behaviors among agents, and we discussed the implementation challenges of reputation systems as a disciplinary mechanism.

An important observation is that trustless cross-ledger swap protocols should use disciplinary mechanisms such as collateral deposit.
The implementation, cost, performance, and complexity of various protocols on permissionless blockchains supporting smart contracts -- e.g. Ethereum, EOS, Tezos \cite{Perez2020e}, and Neo \cite{neo} -- thus merit future research.

\newpage
\appendix
\label{sec:proofs}
\section*{Appendix}


Most of the arguments in the proofs below follow the hypothesis that an agent always takes the actions which maximize their expected utility, taking into account future and possibly adversarial actions from the other agent.
We always describe the key conditions (inequalities) to be verified but provide limited details on the derivations as they can be long and tedious.

\paragraph{Proof of \Cref{pkt:mal}}
At time 2 if $\Tcal_a=l$, then Alice loses $\frac{P_2}{2}$ in utility by playing $c$ instead of $s$.
Similarly, at time 1 if $\Tcal_b=l$, then Bob gets $\frac{P_1}{2}$ in utility by playing $s$ whereas he expects to receive $\cExp{\Ucal(b,l)}{\{c,c\}}=\mu_a(P_1-P_0) + (1-\mu_a)(\frac{P_1}{2}-P_0)$ if he plays $c$.
We have $\cExp{\Ucal(b,l)}{\{c,c\}} < \frac{P_1}{2}$ since $\delta<\frac{P_0}{2}$ and $\mu_a\le1$, hence a malicious Bob plays $s$.

\paragraph{Proof of Corollary~\ref{pkt:pfail}} 
The transaction succeeds only if Alice and Bob are honest which happens with probability $\Pa[\Tcal_a=\Tcal_b=h]=\mu_a\mu_b$.

\paragraph{Proof of \Cref{pkt:bhonest}}
We have $\cExp{\Ucal(b,h)}{\{c,w,c\}} = \mu_a (P_1 - P_0 + \alpha_{b,h}) + (1-\mu_a)(-\alpha_{b,h} + \frac{P_1}{2} - P_0) $ and $\cExp{\Ucal(b,h)}{\{c,w,s\}} = -\alpha_{b,h} + \frac{P_1}{2}$.
We obtain that $\br(b,\{c,w\})=c$ by taking $P_1=P_0-\delta$.

\paragraph{Proof of \Cref{pkt:ahonest}}
We have $\br(h,\{c,w,c,w\})=c$ if and only if $\alpha_{a,h} + P_0 - P_2 > -\alpha_{a,h} + P_0 - \frac{P_2}{2}$ which is equivalent to $\alpha_{a,h}>\frac{P_0+2\delta}{4}$.
Then, with $\br(h,\{c,w,c,w\})=c$, we have that $\cExp{\Ucal(a,h)}{\{c\}}=\mu_b\alpha_{a,h} + (1-\mu_b)(-\alpha_{a,h} - \frac{P_0}{2})$ and $\cExp{\Ucal(a,h)}{\{s\}}= - \alpha_{a,h}$.
Therefore, for agent $a$ to be honest it must also be that $\mu_b > \frac{P_0}{4\alpha_{a,h} + P_0}$

\paragraph{Proof of \Cref{pkt:collateral}}
This is immediate as malicious agents would never be able to make any profit by exiting prematurely the transaction.



\newpage

\bibliographystyle{splncs04}
\bibliography{references}
\end{document}